\documentclass[aps]{revtex4}
\usepackage{graphicx}
\usepackage{dcolumn}
\usepackage{amsmath}
\usepackage[latin1]{inputenc}

\begin{document}
\title{Size-dependence of anisotropic exchange interaction in InAs/GaAs quantum dots}

\author{R. Seguin\footnote{e-mail: {\sf seguin@sol.physik.tu-berlin.de}}, S. Rodt, A. Schliwa, K. P\"otschke, U.~W.~Pohl, and D.~Bimberg} 

\affiliation{Institut f{\"u}r Festk{\"o}rperphysik, Technische Universit{\"a}t Berlin, Hardenbergstr. 36, 10623 Berlin, Germany}

\begin{abstract}
  A comprehensive study of the exchange interaction between charge carriers in self-organized InAs/GaAs quantum dots is presented. Single quantum-dot cathodoluminescence spectra of quantum dots of different sizes are analyzed. Special attention is paid to the energetic structure of the charged excited exciton (hot trion). A varying degree of intermixing within the hot trion states leads to varying degrees of polarization of the corresponding emission lines. The emission characteristics change from circularly polarized for small quantum dots to elliptically polarized for large quantum dots. The findings are explained by a change of magnitude of the anisotropic exchange interaction and compared to the related effect of fine-structure splitting in the neutral exciton and biexciton emission.
\end{abstract}

\maketitle                  

\section{Introduction}

Electron-hole exchange interaction in semiconductor quantum dots has been the subject of a lively debate in recent years. The focus was placed on its impact on the exciton ground state \cite{seguin01, seguin02, seguin03}. In systems with low confinement symmetry ($C_{2v}$) \cite{seguin11, seguin12} the anisotropic exchange interaction lifts the twofold degeneracy of the exciton bright states yielding the so-called exciton fine-structure splitting (FSS). If the FSS is larger than the exciton's homogeneous linewidth, it inhibits the emission of entangled photon pairs \cite{seguin18}. Thus, the FSS is the key parameter determining the suitability of quantum dots for single-photon sources for quantum cryptography \cite{seguin04} and quantum computing \cite{seguin17}. The two bright exciton (X) states yield two spectrally narrow lines at low temperature which are commonly linearly polarized in the [110] and [1$\bar{1}$0] crystal directions, respectively. Since the biexciton (XX) net spin is 0, the FSS can also be observed in the XX to X decay (Fig.\ 1(a)) \cite{seguin05}.

Moreover, exchange interaction alters the energetic structure of the excited positive trion (X+$^{\ast}$) which comprises one singlet and three triplet states all of which are twofold degenerate \cite{seguin06}. Contrary to the X case, this degeneracy is not lifted under exchange interaction. However, it leads to a mixing of the triplet states \cite{seguin06} which can be probed by analyzing the luminescence lines from the positively charged biexciton (XX+) to X+$^{\ast}$ decay. Such probing is possible since the ground state of the initial state, namely the XX+, is completely degenerate, analogous to the XX case. The degree of polarization of these lines reflects the degree of intermixing of the different charged trion states and therefore the magnitude of the anisotropic exchange interaction (Fig.\ 1(b)) \cite{seguin07}. Understanding this effect is crucial for the use of trion states in future spin memory devices \cite{seguin08}.

\begin{figure}[htb]
  \includegraphics[width=.6\textwidth]{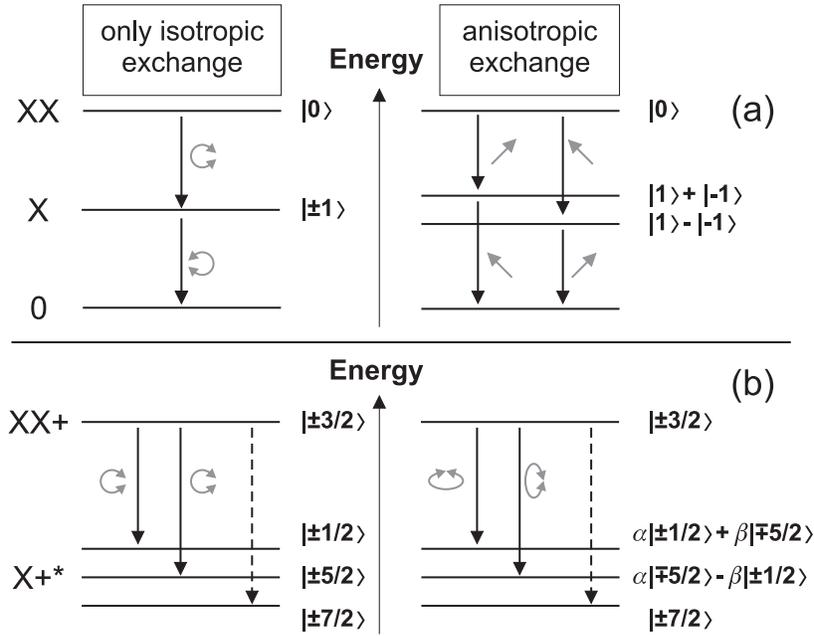}
\caption{Influence of the anisotropic electron-hole exchange interaction on the biexciton$\,\rightarrow\,$exciton$\,\rightarrow\,$0 (a) and the charged-biexciton$\,\rightarrow\,$excited-trion (b) cascades. $\left|M\right\rangle$ gives the total angular momentum of the corresponding excitonic complex. Gray circles, arrows, and ellipses mark the polarization of the corresponding emission. Dashed lines mark forbidden transitions. The X+$^{\ast}$ singlet states are omitted for simplicity.}
\label{fig1}
\end{figure}

We demonstrate here that the size of the quantum dots has a strong impact on the magnitude of the anisotropic exchange interaction by monitoring the degree of polarization of the XX+ to X+$^{\ast}$emission lines and compare our results to a previous study of the excitonic FSS obtained on the same sample \cite{seguin15} .

\section{Experimental}
The sample investigated was grown by metalorganic chemical vapor deposition on a GaAs(001) substrate. A 300~nm thick GaAs buffer layer followed by a 60 nm Al$_{0.6}$Ga$_{0.4}$As diffusion barrier and 90 nm GaAs were grown. For the QD layer nominally 1.9 monolayers of InAs were deposited followed by a 5 s growth interruption. Antimony was added during the growth interruption acting as a surfactant. Subsequently, the QDs were capped with 50 nm of GaAs. Finally, a 20 nm Al$_{0.33}$Ga$_{0.67}$As diffusion barrier and a 10 nm GaAs capping layer were deposited. 

The sample was examined with a JEOL JSM 840 scanning electron microscope equipped with a cathodoluminescence setup described in Ref. \cite{seguin09}. It was mounted onto a He flow cryostat, which provided temperatures as low as 6 K. The luminescence was dispersed by a 0.3 m spectrometer. The detection system consisted of a liquid-nitrogen cooled Si charge-coupled-device camera and a liquid-nitrogen cooled InGaAs diode array. The FWHM as given by the setup was \mbox{$\approx$140 $\mu$eV} at 1.3 eV. Using lineshape analysis we can determine the energetic position of resolution-limited emission lines to an accuracy of $\approx$20 $\mu$eV. Metallic shadow masks with 100 nm apertures were applied in order to reduce the number of simultaneously probed quantum dots.

The quantum dots examined form a series of subensembles, each representing quantum dots with a fixed height, ranging from two to more than nine InAs monolayers \cite{seguin14,seguin16}. In the ensemble spectra, the subensembles emit rather narrow peaks (FWHM $\approx$ 30 meV). They superimpose to give the complete ensemble spectrum, consisting of eight of such peaks. This allows for a precise assignment of a given emission energy to a corresponding quantum dot height.

\begin{figure}[htb]
  \includegraphics[width=.5\textwidth]{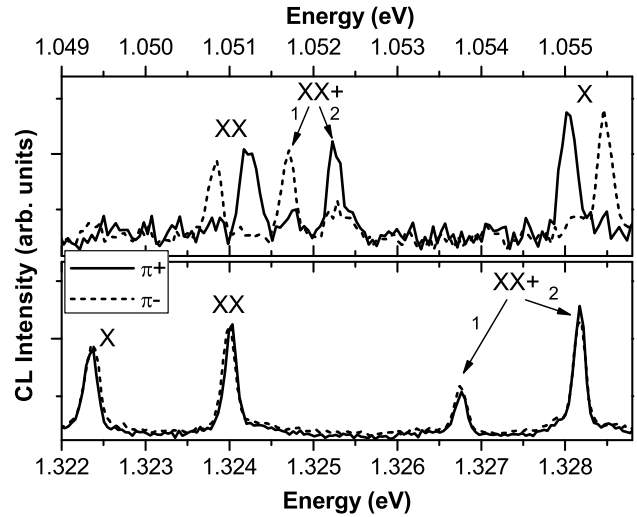}
\caption{Linearly polarized spectra for two different quantum dots emitting at high and low energies are shown. The different FSS magnitudes and the different XX+ degrees of polarization are clearly visible. Note that both X-axes are drawn to the same scale for easier comparison.}
\label{fig2}
\end{figure}

Two spectra of a large and a small single quantum dot emitting at low and high energies respectively are shown in Fig.\ 2. The different excitonic complexes were identified according to Ref.\ \cite{seguin10}. While the XX+ emission lines from the small quantum dot are circularly polarized the large quantum dot emits elliptically polarized XX+ lines. Likewise, the FSS for the small quantum dot is considerably smaller than for the large quantum dot.

In order to examine the influence of quantum-dot height on the magnitude of anisotropic exchange interaction in detail, a number of quantum dots were probed with transition energies ranging from 1.05 to 1.35 eV. 

The degree of polarization $p$ of the XX+ lines was determined by measuring their intensities for both polarization directions and using

\begin{equation}
p = \frac{I_{\pi +}-I_{\pi -}}{I_{\pi +}+I_{\pi -}}.
\end{equation}

The FSS values were determined by taking the average of the energetic differences of the two polarized X and the two polarized XX lines. The FSS is defined to be positive, if the X line at lower energy is polarized along the [1$\bar{1}$0] crystal direction following our previous work \cite{seguin15}.

The results are displayed in Fig.\ 3.  Note that only absolute values of polarization degrees are shown in Fig.\ 3(a) for simplicity. The preferential polarization axis for the XX+ line at lower (higher) energy is always along the [110] ([1$\bar{1}$0]) crystal direction. The FSS values range from -80 $\mu$eV to 520 $\mu$eV \cite{seguin15}.

\begin{figure}[ht]
  \includegraphics[width=.9\textwidth]{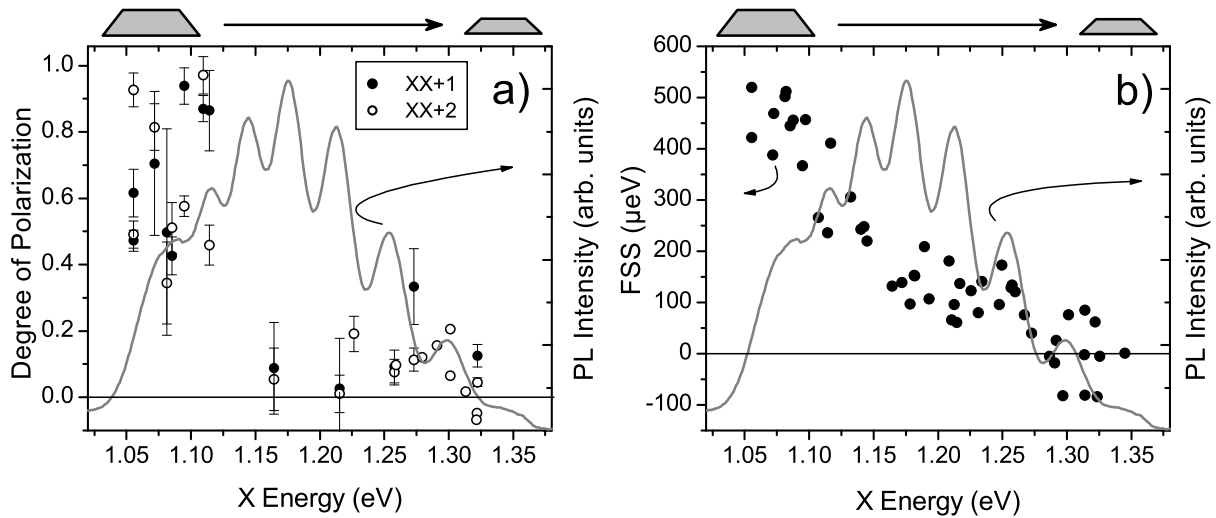}
\caption{a) shows the degree of polarization for the two XX+ emission lines versus the corresponding exciton transition energy. b) shows the measured FSS values taken from Ref \cite{seguin15}. The gray line displays the ensemble luminescence spectrum with its pronounced modulation.}
\label{fig3}
\end{figure}

The measurements clearly show that the anisotropic part of electron-hole exchange is a function of quantum dot size. 

Anisotropic electron-hole exchange in quantum dots a\-ri\-ses, when the the symmetry of the confining potential is lower than $C_{4v}$. Sources for such symmetry lowering include structural anisotropy of the quantum dots, strain-induced piezoelectricity \cite{seguin11, seguin12}, and atomistic symmetry anisotropy \cite{seguin13}. Structural elongation of the quantum dots however can be ruled out to be the main source for the observed exchange effects. TEM measurements of our quantum dots show no significant anisotropy \cite{seguin14} and numerical modeling fails to reproduce experimental results \cite{seguin15}. Piezoelectricity provides a possible explanation for the observed trend \cite{seguin15}. Its magnitude is proportional to the occurring shear strain in the QDs. Due to the lattice mismatch between GaAs and InAs, the shear strain is larger for larger QDs \cite{seguin11}. Hence small (big) QDs have weak (strong) shear strain components leading to weak (strong) piezoelectric fields and consequently to small (large) values of the fine-structure splitting and the XX+ lines show small (large) degrees of polarization. The importance of strain for the magnitude of the FSS has been noted before \cite{seguin19}. The role of atomistic symmetry anisotropy is not assessed in detail yet. Therefore it cannot be excluded when drawing a complete picture of exchange interaction in quantum dots.

\section{Conclusion}
While the number of participating particles varies between two (excitons) and three (trions), the underlying physical effect leading to the FSS and a finite polarization degree of the XX+ lines is the same, namely the anisotropic exchange interaction. We have shown that the magnitude of the excitonic FSS and the polarization degree of the charged XX lines are indeed correlated. They both increase with decreasing exciton recombination energy and thus increasing quantum dot size, clearly indicating a direct dependence of the magnitude of the anisotropic exchange interaction on QD size.

\section{Acknowledgement}
  This work has been supported in part by the Deutsche For\-schungs\-ge\-mein\-schaft through SFB 296 and the SANDiE Network of Excellence of the European Commission, Contract No. NMP4-CT-2004-500101.

\end{document}